\documentclass[11pt,a4paper]{article}
\usepackage[hyperref]{emnlp2018}
\usepackage{times}
\usepackage{latexsym}
\usepackage{url}

\usepackage{helvet}
\usepackage{courier}
\usepackage{graphicx}

\usepackage{adjustbox}
\usepackage{amsmath,amssymb,amsthm}
\usepackage{mathabx}
\usepackage{multirow}
\usepackage{url}
\usepackage{hhline}

\usepackage{enumerate}

\usepackage{tikz,pgfplots}

\usepackage{pgfplots}
\usepackage{xcolor}
\usetikzlibrary{arrows,intersections}

\usepackage{multirow}
\usepackage{pgfplots}

\usepackage{algorithm}
\usepackage[noend]{algpseudocode}

\usepackage{subcaption}
\usepackage{multirow}
\include{plots}
\definecolor{bblue}{HTML}{4F81BD}
\definecolor{rred}{HTML}{C0504D}
\definecolor{ggreen}{HTML}{9BBB59}
\definecolor{ppurple}{HTML}{9F4C7C}
\definecolor{Dark scarlet}{HTML}{560319}
\definecolor{Forest green}{HTML}{1E4D2B}

\usepackage[font=small]{caption}
\DeclareCaptionFont{10pt}{\fontsize{10pt}{12pt}\selectfont}
\aclfinalcopy 

\title{Document Informed Neural Autoregressive Topic Models
}

\newcommand*{\affaddr}[1]{#1} 
\newcommand*{\affmark}[1][*]{\textsuperscript{#1}}

\author{Pankaj Gupta\affmark[1,2], Florian Buettner\affmark[1], Hinrich Sch\"{u}tze\affmark[2]\\ 
 \affaddr{\affmark[1]Corporate Technology, Machine-Intelligence (MIC-DE), Siemens AG  Munich, Germany}\\
  \affaddr{\affmark[2]CIS, University of Munich (LMU) Munich, Germany} \\
  {\tt \{ pankaj.gupta, buettner.florian\}@siemens.com}\\
  {\tt pankaj.gupta@campus.lmu.de |  inquiries@cislmu.org}
}

\date{}

\begin{document}
\maketitle

\begin{abstract}
Context information around words helps in determining their actual meaning, for example ``networks" used 
in contexts of {\it artificial neural networks} or {\it biological neuron networks}. 
Generative topic models infer topic-word distributions, taking no or only little context into account. 
Here, we extend a neural autoregressive topic model to exploit the full context information around words in a document in a language modeling fashion. 
This results in an improved performance in terms of generalization, 
interpretability and applicability. 
We apply our modeling approach to seven data sets from various domains and demonstrate  
that our approach consistently outperforms state-of-the-art generative topic models. 
With the learned representations, we show on an average a gain of $9.6$\% ($0.57$ Vs $0.52$) in precision 
at retrieval fraction $0.02$ and $7.2$\% ($0.582$ Vs $0.543$) in $F1$ for text categorization. 
\end{abstract}

\section{Introduction}

Probabilistic topic models, such as LDA \cite{Blei:81},  Replicated Softmax (RSM) \cite{Salakhutdinov:82} and  Document Autoregressive 
Neural Distribution Estimator (DocNADE) 
 \cite{Hugo:82} are often used to extract 
topics from text collections and learn 
document representations to perform NLP tasks such 
as information retrieval (IR), document classification or summarization.

To motivate our task, assume that we conduct topic analysis on a collection of research papers from 
NIPS conference, where
one of the popular terms is ``networks". 
However, without 
context information (nearby and/or distant words), its actual meaning is ambiguous since it can refer to such different concepts as 
{\it artificial neural networks} in {\it computer science}  or {\it biological neural networks} in {\it neuroscience} or  {\it Computer}/{\it data networks} in {\it telecommunications}. 
Given the 
context, one can determine the actual meaning of  ``networks", for instance,  
``Extracting rules from artificial neural \underline{networks} with distributed representations", or 
``Spikes from the presynaptic neurons and postsynaptic neurons in small \underline{networks}" or 
``Studies of neurons or \underline{networks} under noise 
in artificial neural \underline{networks}" 
or ``Packet Routing in Dynamically Changing  \underline{Networks}". 
%

\begin{figure*}[t]
  \centering
  \includegraphics[scale=.73]{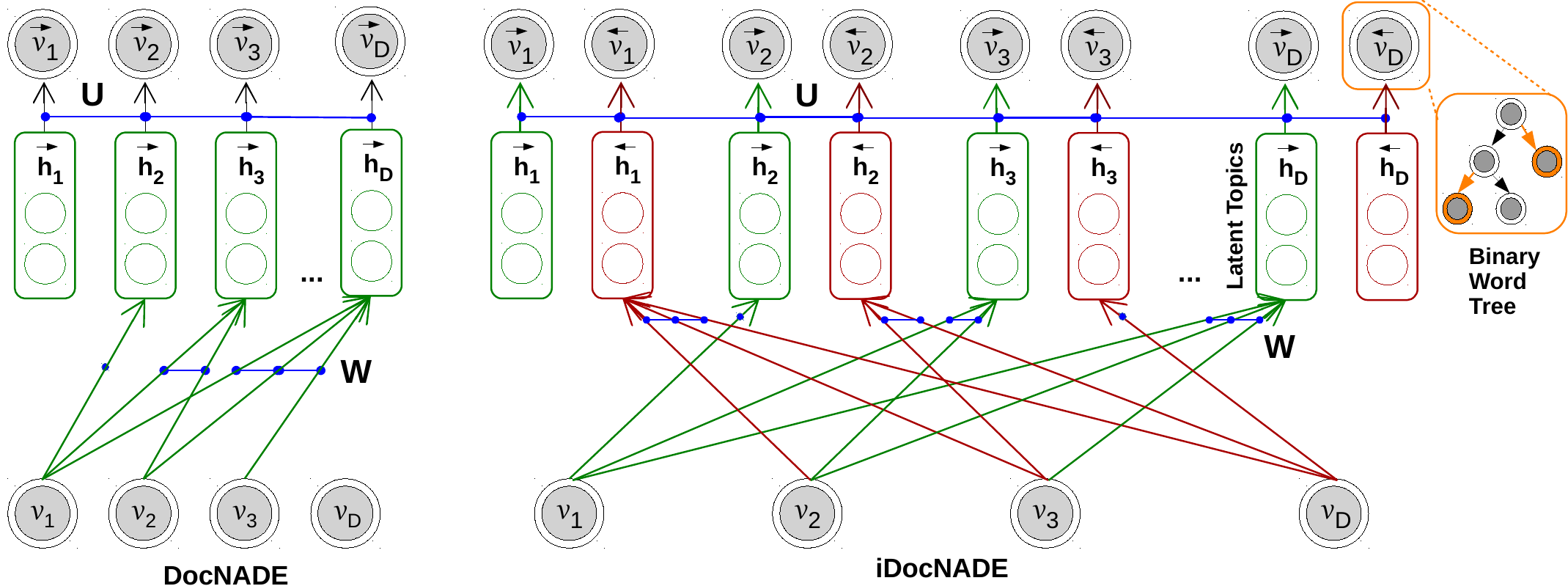}
  \caption{{\bf (Left):} {\it DocNADE}  {\bf (Right):} {\it iDocNADE} (the {\it proposed} model).  
Blue colored lines signify the connections that share parameters.  
The observations ({\it double circle}) for each word $v_i$ are multinomial. 
Hidden vectors in {\it green} and {\it red} colors identify the forward and backward network layers, respectively.   
Symbols $\overrightarrow{v}_{i}$ and $\overleftarrow{v}_{i}$ 
represent the autoregressive conditionals $p(v_i | {\bf v}_{< i})$ and $p(v_i | {\bf v}_{> i})$, respectively. 
Connections between each $v_i$ and hidden units are  shared, 
and each conditional {$\overrightarrow{v}_{i}$}  (or {$\overleftarrow{v}_{i}$}) is decomposed into a tree of binary logistic regressions, i.e. hierarchical softmax. Best viewed in color.
}
  \label{fig:AutoregressiveNetworks}
\end{figure*}

Generative topic models such as LDA or DocNADE infer topic-word distributions that can be used to estimate a document likelihood. 
While basic models such as LDA do not account for context information when inferring these distributions, more recent approaches
such as DocNADE achieve {\it amplified word and document  likelihoods} by accounting for words preceding a word of interest in a document.
More specifically, DocNADE \cite{Hugo:82,Hugo:83} (Figure \ref{fig:AutoregressiveNetworks}, Left) is a probabilistic graphical model that learns topics over sequences of words, corresponding to a language model \cite{Manning:82,Bengio:82} 
that can be interpreted as a neural network with several parallel hidden layers. 
To predict the word $v_i$, each hidden layer ${\bf h}_i$ takes as input the sequence of preceding words ${\bf v}_{<i}$.  
However, it does {\it not} take into account the following words ${\bf v}_{>i}$ in the sequence. 
Inspired by bidirectional language models \cite{Mousa:82} and recurrent neural networks \cite{elman1990finding,Gupta:89,Gupta:82,
Vu:82,Gupta:87}, trained to
predict a word (or label) depending on its full left and right contexts, we extend DocNADE and incorporate 
full contextual information (all words around ${v}_i$) at each hidden layer ${\bf h}_i$ when predicting the word ${v}_i$ in a language modeling fashion with neural topic modeling.

{\bf Contribution:}
{\bf (1)} We propose an advancement in neural autoregressive topic model by incorporating full 
contextual information around words in a document to boost the likelihood of each word (and document). 
We demonstrate using 7 data sets from various domains that this enables learning better ({\it informed}) document representations in terms of {\it generalization} (perplexity), {\it interpretability} (topic coherence)  and 
{\it applicability} (document retrieval and classification).  
We name the proposed topic model as {\it Document Informed Neural Autoregressive Distribution Estimator} ({\bf iDocNADE}). 

{\bf (2)} With the learned representations, we show a gain of $9.6$\% ($.57$ Vs $.52$) in precision at retrieval fraction 0.02 and $7.2$\% ($.582$ Vs $.543$) in $F1$ for text categorization compared to the DocNADE model (on average over 6 data sets).   

The {\it code} and pre-processed data 
is available at {\tt https://github.com/pgcool/iDocNADE}\footnote{we will release soon upon acceptance}.%


\section{Neural Autoregressive Topic Models}
RSM \cite{Salakhutdinov:82,Gupta:85}, a probabilistic undirected topic model is a generalization of  the energy-based Restricted Boltzmann
Machines RBM \cite{Hinton:83,Gupta:86,Gupta:91}, 
which can be used to model word counts. 
NADE \cite{Hugo:84} decomposes the joint distribution of observations into autoregressive conditional distributions, 
modeled using non-linear  functions.  
Unlike for RBM and RSM, this leads to tractable 
gradients of the 
data negative 
log-likelihood but can only be used to model binary observations.

{\bf DocNADE} (Figure \ref{fig:AutoregressiveNetworks}, Left), a generative neural autoregressive topic model to 
account for word counts, is inspired by RSM and NADE. 
For a document ${\bf v}=[v_1, ..., v_D]$ of size $D$, it models the joint distribution $p({\bf v})$ of 
all words $v_i$, where $v_i \in \{1,..., K\}$ 
is the index of the $i$th word in the dictionary of vocabulary size $K$. This is achieved by decomposing it as a product of  
conditional distributions i.e. $p({\bf v})= \prod_{i=1}^{D} p(v_i | {\bf v}_{<i})$ 
and computing each autoregressive conditional  $p(v_i | {\bf v}_{<i})$  via a feed-forward neural network for $i \in \{1,...D\}$,\vspace{-0.1cm}   
\begin{align*}
\begin{split}
\overrightarrow{\bf h}_i({\bf v}_{<i}) & =  g (D{\bf c} + \sum_{k<i} {\bf W}_{:, v_k})
\end{split}\\
\begin{split}
p (v_i = w | {\bf v}_{<i}) & = \frac{\exp (b_w + {\bf U}_{w,:} \overrightarrow{\bf h}_i ({\bf v}_{<i}))}{\sum_{w'} \exp (b_{w'} + {\bf U}_{w',:} \overrightarrow{\bf h}_i ({\bf v}_{<i}))}
\end{split}
\end{align*}
where ${\bf v}_{<i} \in \{v_1, ...,v_{i-1}\}$.  
$g(\cdot)$ is a non-linear activation function, ${\bf W} \in \mathbb{R}^{H \times K}$ and ${\bf U} \in \mathbb{R}^{K \times H}$ 
are weight 
matrices, ${\bf c} \in \mathbb{R}^H$ and ${\bf b} \in \mathbb{R}^K$ are bias parameter 
vectors. $H$ is the number of hidden units (topics). 
${\bf W}_{:,<i}$ is a matrix made of the $i-1$ first columns of ${\bf W}$. Therefore, 
the log-likelihood $\mathcal{L}$ of a document {\bf v} in DocNADE is computed as:\vspace{-0.3cm}
\begin{equation}\label{eq:DocNADEloglikelihood}
\mathcal{L}^{DocNADE}({\bf v})  =  \sum_{i=1}^{D} \log p (v_i | {\bf v}_{<i})
\end{equation}

The probability of the word $v_i$ 
is computed using a position-dependent hidden layer $\overrightarrow{\bf h}_i({\bf v}_{<i})$ 
that learns a representation based on all previous words ${\bf v}_{<i}$; 
however it does {\it not} incorporate the following words ${\bf v}_{>i}$. Taken together, the likelihood $p(\bf v)$ of any document of arbitrary length can be computed. 
Note that, following RSM, we re-introduced the scaling factor $D$ in computing ${\bf h}_i$ 
to account for documents of different lengths, that is ignored in the original DocNADE formulation.

{\bf iDocNADE} (Figure \ref{fig:AutoregressiveNetworks}, Right), our {\it proposed} model accounts for the full context information 
(both previous ${\bf v}_{<i}$ and following ${\bf v}_{>i}$ words) around each word $v_i$  for a document ${\bf v}$. 
Therefore, the log-likelihood $\mathcal{L}^{iDocNADE}$ for a document $\bf v$ in {\it iDocNADE} is computed using forward and backward language models as:\vspace{-0.2cm}
\begin{align}
\begin{split}
\log p({\bf v})  = \frac{1}{2}  \sum_{i=1}^{D}  \underbrace{\log p(v_i | {\bf v}_{<i})}_{\mbox{forward}} & +   \underbrace{\log p(v_i | {\bf v}_{>i})}_{\mbox{backward}}  
\end{split}\\ \label{eq:iDocNADEloglikelihood}
\begin{split}
\mathcal{L}^{iDocNADE}({\bf v}) =  \frac{1}{2} \Big[ \overrightarrow{\mathcal{L}} ({\bf v}) & + \overleftarrow{\mathcal{L}} ({\bf v})  \Big]
\end{split}
\end{align}
where $\mathcal{L}^{iDocNADE}$ is the mean of the forward ($\overrightarrow{\mathcal{L}}$) and backward ($\overleftarrow{\mathcal{L}}$)  log-likelihoods. 
This is achieved in a bi-directional language modeling 
and feed-forward fashion by computing  position dependent {\it forward} ($\overrightarrow{\bf h}_i$) 
and {\it backward} ($\overleftarrow{\bf h}_i$) hidden layers   
for each word $i$, as:\vspace{-0.1cm} 
\begin{align*}
\begin{split}
\overrightarrow{\bf h}_i({\bf v}_{<i}) =  g (D \overrightarrow{\bf c} +\sum_{k<i} {\bf W}_{:, v_k}) 
\end{split}\\
\begin{split}
\overleftarrow{\bf h}_i({\bf v}_{>i}) =  g (D \overleftarrow{\bf c} + \sum_{k>i} {\bf W}_{:, v_k}) 
\end{split}
\end{align*}

where $\overrightarrow{\bf c} \in \mathbb{R}^H$ and $\overleftarrow{\bf c} \in \mathbb{R}^H$ are bias parameters 
in forward and backward passes, respectively. $H$ is the number of hidden units (topics). 

Two autoregressive conditionals are computed for each $i$th word using the forward and backward hidden vectors, as:\vspace{-0.3cm}  
\begin{align*}
\begin{split}
p (v_i = w | {\bf v}_{<i}) = \frac{\exp (\overrightarrow{b}_w + {\bf U}_{w,:} \overrightarrow{\bf h}_i ({\bf v}_{<i}))}{\sum_{w'} \exp (\overrightarrow{b}_{w'} + {\bf U}_{w',:} \overrightarrow{\bf h}_i ({\bf v}_{<i}))}
\end{split}\\
\begin{split}
p (v_i = w | {\bf v}_{>i}) = \frac{\exp (\overleftarrow{b}_w + {\bf U}_{w,:} \overleftarrow{\bf h}_i ({\bf v}_{>i}))}{\sum_{w'} \exp (\overleftarrow{b}_{w'} + {\bf U}_{w',:} \overleftarrow{\bf h}_i ({\bf v}_{>i}))}
\end{split}
\end{align*}
for  $i \in [1, ..., D]$ where  $\overrightarrow{\bf b} \in \mathbb{R}^K$ and $\overleftarrow{\bf b} \in \mathbb{R}^K$ are biases in forward and 
backward passes, respectively.  Observe, the parameters {\bf W} and {\bf U} are shared in the forward and backward networks.
\begin{algorithm}[t]
\caption{{\small Computation of $\log p({\bf v})$ using iDocNADE}}\label{trainingiDocNADE}
\begin{algorithmic}[1]
\Statex \textbf{Input}: A training document vector {\bf v}
\Statex \textbf{Parameters}: \{\overrightarrow{\bf b},  \overleftarrow{\bf b}, \overrightarrow{\bf c}, \overleftarrow{\bf c}, {\bf W}, {\bf U}\}
\Statex \textbf{Output}: $\log p({\bf v})$
\State $\overrightarrow{\bf a} \gets \overrightarrow{\bf c}$  
\State  $\overleftarrow{\bf a} \gets \overleftarrow{\bf c} +  \sum_{i >1}{\bf W}_{:, v_{i}}$
\State $ q({\bf v}) = 1$
\For{$i$ from $1$ to $D$}
        \State $\overrightarrow{\bf h}_{i}  \gets g(\overrightarrow{\bf a})$
        \State $\overleftarrow{\bf h}_{i}  \gets g(\overleftarrow{\bf a})$
        \State $  p(v_{i} | {\bf v}_{<i}) = 1$
        \State $  p(v_{i} | {\bf v}_{>i}) = 1$
        \For{$m$ from $1$ to $| {\bf \pi}(v_{i})|$}
                   \State $  p(v_{i} | {\bf v}_{<i}) \gets   p(v_{i} | {\bf v}_{<i})    p({\bf \pi}(v_{i})_m| {\bf v}_{<i})$
                   \State $  p(v_{i} | {\bf v}_{>i}) \gets   p(v_{i} | {\bf v}_{>i})    p({\bf \pi}(v_{i})_m| {\bf v}_{>i})$
        \EndFor
       \State $ q({\bf v}) \gets  q({\bf v})   p(v_{i} | {\bf v}_{<i})  p(v_{i} | {\bf v}_{>i})$
      \State $\overrightarrow{\bf a} \gets \overrightarrow{\bf a} + {\bf W}_{:, v_{i}}$ 
      \State $\overleftarrow{\bf a} \gets \overleftarrow{\bf a} - {\bf W}_{:, v_{i}}$ 
\EndFor
\State $\log p({\bf v}) \gets \frac{1}{2} \log q({\bf v})$
\end{algorithmic}
\end{algorithm}

\begin{algorithm}[t]
\caption{{Computing gradients of $-\log p({\bf v})$ in iDocNADE}}\label{gradientiDocNADE}
\small{ 
\begin{algorithmic}[1]
\Statex \textbf{Input}: A training document vector {\bf v}
\Statex \textbf{Parameters}: \{\overrightarrow{\bf b},  \overleftarrow{\bf b}, \overrightarrow{\bf c}, \overleftarrow{\bf c}, {\bf W}, {\bf U}\}
\Statex \textbf{Output}: $\delta\overrightarrow{\bf b},  \delta\overleftarrow{\bf b}, \delta\overrightarrow{\bf c}, \delta\overleftarrow{\bf c}, \delta{\bf W}, \delta{\bf U}$ 
\State $\overrightarrow{\bf a} \gets 0$; $\overleftarrow{\bf a} \gets 0$; $\overrightarrow{\bf c} \gets 0$;  $\overleftarrow{\bf c} \gets 0$; $\overrightarrow{\bf b} \gets 0$;  $\overleftarrow{\bf b} \gets 0$ 
\For{$i$ from $D$ to $1$}
        \State $\delta\overrightarrow{\bf h}_i \gets 0$
         \State $\delta\overleftarrow{\bf h}_i \gets 0$
        \For{$m$ from $1$ to $| {\bf \pi}(v_{i})|$}
               \State $\overrightarrow{b}_{l{(v_i)_m}} \gets \overrightarrow{b}_{l{(v_i)_m}} + ( p(\pi(v_i)_m | {\bf v}_{<i}) - \pi(v_i)_m)$
               \State $\overleftarrow{b}_{l{(v_i)_m}} \gets \overleftarrow{b}_{l{(v_i)_m}} + ( p(\pi(v_i)_m | {\bf v}_{>i}) - \pi(v_i)_m)$
               \State $\delta\overrightarrow{\bf h}_i \gets \delta\overrightarrow{\bf h}_i  + (p( \pi(v_i)_m | {\bf v}_{<i}) -  \pi(v_i)_m) {\bf U}_{l(v_i)m,:}$               
               \State $\delta\overleftarrow{\bf h}_i \gets \delta\overleftarrow{\bf h}_i  + (p( \pi(v_i)_m | {\bf v}_{>i}) -  \pi(v_i)_m) {\bf U}_{l(v_i)m,:}$
              \State $\delta{\bf U}_{l(v_i)_m}  \gets \delta{\bf U}_{l(v_i)_m} + ( p(\pi(v_i)_m | {\bf v}_{<i}) - \pi(v_i)_m) \overrightarrow{\bf h}_i^{T} +  (p(\pi(v_i)_m | {\bf v}_{>i}) - \pi(v_i)_m) \overleftarrow{\bf h}_i^{T}$
        \EndFor
        \State $\delta\overrightarrow{\bf g} \gets \overrightarrow{\bf h}_i  \circ (1 - \overrightarrow{\bf h}_i)$ \# for sigmoid activation
        \State $\delta\overleftarrow{\bf g} \gets \overleftarrow{\bf h}_i  \circ (1 - \overleftarrow{\bf h}_i)$ \# for sigmoid  activation
        \State $\delta\overrightarrow{\bf c} \gets  \delta\overrightarrow{\bf c} + \delta\overrightarrow{\bf h}_i  \circ \delta\overrightarrow{\bf g}$
        \State $\delta\overleftarrow{\bf c} \gets  \delta\overleftarrow{\bf c} + \delta\overleftarrow{\bf h}_i  \circ \delta\overleftarrow{\bf g}$
        \State $ \delta{\bf W}_{:,v_i} \gets  \delta{\bf W}_{:,v_i} + \delta\overrightarrow{\bf a} +  \delta\overleftarrow{\bf a}$
        \State $\delta\overrightarrow{\bf a} \gets \delta\overrightarrow{\bf a} +  \delta\overrightarrow{\bf h}_i  \circ \delta\overrightarrow{\bf g}$
        \State $\delta\overleftarrow{\bf a} \gets \delta\overleftarrow{\bf a} +  \delta\overleftarrow{\bf h}_i  \circ \delta\overleftarrow{\bf g}$
\EndFor
\end{algorithmic}}
\end{algorithm}

{\bf Learning}: 
Similar to DocNADE, the autoregressive conditionals  $p (v_i = w | {\bf v}_{<i})$ and $p (v_i = w | {\bf v}_{>i})$ 
in iDocNADE are computed by a neural network for each word $v_i$, allowing efficient learning of {\it informed} 
representations $\overrightarrow{\bf h}_{i}$ and  $\overleftarrow{\bf h}_{i}$, 
as it consists simply of a linear transformation followed by a  non-linearity. 
Observe that the weight $\bf W$ is the same across all conditionals and ties (blue colored lines) contextual observables by computing each $\overrightarrow{\bf h}_{i}$ 
or $\overleftarrow{\bf h}_{i}$.   

{\it Binary word tree to compute conditionals}: 
The computations of each of the autoregressive conditionals  $p (v_i = w | {\bf v}_{<i})$ and $p (v_i = w | {\bf v}_{>i})$  
require time linear in $K$, which is expensive to compute for $i \in [1, 2, ...D]$.  
Following \newcite{Hugo:82}, we decompose the computation of the conditionals to achieve 
a complexity logarithmic in $K$. All words in the documents are randomly assigned to a different leaf in a binary tree 
and the probability of a word is computed as the probability of reaching its associated leaf from the root. 
Each left/right transition probability is modeled using a binary logistic regressor  with 
the hidden layer $\overrightarrow{\bf h}_{i}$ or  $\overleftarrow{\bf h}_{i}$ as its input.  
In the binary tree, the probability of a given word is computed by multiplying each of the left/right 
transition probabilities  along the tree path.  

Algorithm \ref{trainingiDocNADE} shows the computation of $\log p({\bf v})$ using the iDocNADE structure, where the autogressive conditionals (lines 10 and 11) for each word $v_i$ are obtained from the forward and backward networks and modeled into a binary word tree, where $\pi (v_{i})$ denotes the sequence of 
binary left/right choices at the internal nodes along the tree path and ${\bf l}(v_i)$ the sequence of tree nodes on that tree path. 
For instance, $l (v_i)_1$ will always be the root of the binary tree and $\pi (v_i)_1$ will be 0 if the word leaf $v_i$ is in the left subtree or 1 otherwise.   
Therefore, each of the forward and backward conditionals are computed as:\vspace{-0.4cm}
\begin{align*}
\begin{split}
p(v_i = w | {\bf v}_{<i}) & = \prod_{m=1}^{|\pi (v_i)|} p(\pi (v_i)_m | {\bf v}_{<i})
\end{split}\\
\begin{split}
p(v_i = w | {\bf v}_{>i}) & = \prod_{m=1}^{|\pi (v_i)|} p(\pi (v_i)_m | {\bf v}_{>i})
\end{split}\\
\begin{split}
p(\pi (v_i)_m | {\bf v}_{<i}) = & g( \overrightarrow{b}_{l{(v_i)_m}} + {\bf U}_{l{(v_i)_m}, :}  \overrightarrow{\bf h}({\bf v}_{<i})) 
\end{split}\\
\begin{split}
p(\pi (v_i)_m | {\bf v}_{>i})  = & g( \overleftarrow{b}_{l{(v_i)_m}} + {\bf U}_{l{(v_i)_m}, :}  \overleftarrow{\bf h}({\bf v}_{>i})) 
\end{split}
\end{align*}
where  ${\bf U} \in \mathbb{R}^{T \times H}$ is the matrix of logistic regressions weights, $T$ is the number of internal nodes in binary tree, 
and  \overrightarrow{\bf b} and \overleftarrow{\bf b} are bias vectors. 

Each of the forward and backward conditionals $p (v_i = w | {\bf v}_{<i})$ or $p (v_i = w |  {\bf v}_{>i} )$  requires the computation of its own hidden layers  $\overrightarrow{\bf h}_i ({\bf v}_{<i})$ and $\overleftarrow{\bf h}_i ({\bf v}_{>i})$, respectively. 
With $H$ being the size of each hidden layer(s) and $D$ the size of the document ${\bf v}$ computing a single layer requires $O(HD)$, and since there are $D$ hidden layers to compute, a naive approach for computing all hidden layers would be in $O(D^2H)$. However, since the weights in the the matrix $\bf W$ are tied, the linear transformations/activations $\overrightarrow{\bf a}$ and $\overleftarrow{\bf a}$ (algorithm \ref{trainingiDocNADE}) can be re-used in every hidden layer and computational complexity reduces to $O(HD)$.  

With the trained {\it iDocNADE} model, the representation ($\overleftrightarrow{\bf \mathfrak{h}} \in { \mathbb R}^{H}$) for a new document {\bf v}*  of size $D^*$ is extracted by summing 
the hidden representations from the forward and backward networks to account for the  
context information around each word in the words' sequence, as 
\begin{align}
\begin{split}
\overrightarrow{\bf h} ({\bf v}^*) =&  g (D^* \overrightarrow{\bf c} +\sum_{k<D^*} {\bf W}_{:, v_k^*}) 
\end{split}\\
\begin{split}
\overleftarrow{\bf h}({\bf v}^*) =&  g (D^* \overleftarrow{\bf c} +\sum_{k>1} {\bf W}_{:, v_k^*})
\end{split}\\
\label{eq:documentrepiDocNADE}
\begin{split}
\mbox{ Therefore;} \  \overleftrightarrow{\bf \mathfrak{h}}  =& \overrightarrow{\bf h}(\bf v^*) + \overleftarrow{\bf h}(\bf v^*)
\end{split}   
\end{align}

The parameters \{\overrightarrow{\bf b},  \overleftarrow{\bf b}, \overrightarrow{\bf c}, \overleftarrow{\bf c}, {\bf W}, {\bf U}\} are learned by minimizing the average negative log-likelihood of the training documents using stochastic gradient descent, as shown in algorithm \ref{gradientiDocNADE}. 
In our proposed formulation of iDocNADE (Figure \ref{fig:AutoregressiveNetworks}), we perform exact inference by computing $\mathcal{L}^{iDocNADE}({\bf v})$ (eqn. \ref{eq:iDocNADEloglikelihood})
as mean of the full forward and backward log likelihoods.  To speed up computations, we can investigate computing a pseudo-likelihood that is further detailed in the {\it supplementary material} (section \ref{pseudolikelihood}). 


\begin{table*}[t]
\centering
\renewcommand*{\arraystretch}{1.2}
\resizebox{0.99\textwidth}{!}{
\begin{tabular}{c||rrr|c|c|c|c||cc|cc||c|c}
\hline
\multirow{3}{*}{\bf Data}       &  \multicolumn{3}{c|}{\bf  Number of Documents} &    \multirow{3}{*}{$\bf K$}   & \multirow{3}{*}{$\bf |C|$}   & \multirow{3}{*}{\bf Label}   & \multirow{3}{*}{\bf Domain} & \multicolumn{4}{c||}{\bf PPL} & \multicolumn{2}{c}{\bf IR-precision ($0.02$)} \\ \cline{2-4} \cline{9-12}   

    &     \multirow{2}{*}{\it train}     &     \multirow{2}{*}{\it dev}   &        \multirow{2}{*}{\it test}    &    &    &    &  &  \multicolumn{2}{c|}{\it DocNADE} & 
\multicolumn{2}{c||}{\it iDocNADE}  & \multicolumn{2}{c}{$T200$} \\ \cline{13-14}

&        &       &         &    &    &    &   & $T50$ &  $T200$ & $T50$ &  $T200$ & {\it DocNADE} &  {\it iDocNADE} \\ \hline\hline

{\it  NIPS}                      &    1,590            &    100           &      50           &    13,649            &   -           &  no           &  Scholar      &  2478   &  2205   &2284    &  \underline{2064}  &-  & -                   \\   \hline

{\it  TREC}       &   5,402            &      50            &   500              &    2,000              &   6           & single            &  Q\&A     &  42    &   42         &    39      &  \underline{39}
&  0.48  &  {\bf 0.55}    \\   \hline 
 
{\it  Reuters8}       &   5,435             &      50            &   2,189              &    2,000              &   8           & single            &  News      &   178 &   172   &   162     &  \underline{152}
& 0.88  &  {\bf 0.89}    \\   \hline  

{\it  Reuters21758}       &   7,769             &      50            &   3,019              &    2,000              &   90           & multi            &  News      &  226        &    215      &  198   & \underline{179} & 0.70  &  {\bf 0.74}     \\   \hline

{\it  Polarity}       &   8,479             &      50            &   2,133              &    2,000              &   2           & single            &  Sentiment    & 310            &    311       &    294    & \underline{292} &  0.51  &  {\bf 0.54}      \\   \hline  

{\it 20NewsGroups}        &  11,337              &   50             &     7,544            &    2,000            &  20            & single            &  News   & 864          &     830          &             836 &  \underline{812} &  0.27  &  {\bf 0.33}       \\   \hline
{\it SiROBs}                   &   27,013             &1,000            &   10,578              &    3,000             &   22           & multi          &  Industry    & 449    &       398        &             392 & \underline{351}
& 0.31  &  {\bf 0.35}   \\   \hline    \hline   
 {\it Average}                 &          \multicolumn{7}{c||}{}          &        650        &     596            &      601          &     \underline{556}        &    0.52 &  {\bf 0.57}  

\end{tabular}}
\caption{Data Statistics and State-of-the-art Comparison ({\it PPL} and {\it IR-precision}) over 50 ($T50$) and 200 ($T200$)  topics for DocNADE and iDocNADE on the six data sets of different domains. 
The \underline{\it underline} and {\bf bold} numbers indicate the best scores in PPL and retrieval task at 0.02 fraction, respectively by iDocNADE.  
See \newcite{Hugo:82} for LDA \cite{Blei:81} performance in terms of PPL, where DocNADE outperforms LDA.  $K$: dictionary size; $C$: class labels.}
\label{datastatistics}
\end{table*}

\section{Evaluation}
We perform quantitative and qualitative evaluations on datasets of varying size with single/multi-class labeled documents from public as well as industrial corpora.
We first demonstrate the generalization capabilities of our proposed model and then the applicability of its representation learning via document retrieval and classification tasks. 

\subsection{Datasets}
We use seven different datasets: 
(1) \texttt{NIPS}: collection of scientific articles from {\url{psiexp.ss.uci.edu/research/programs_data/toolbox.htm} and \url{psiexp.ss.uci.edu/research/programs_data/importworddoccounts.html}}.  
(2) \texttt{TREC}: a set of questions \cite{li2002learning}.   
(3) \texttt{Reuters8}: a collection of news stories, processed and released by \newcite{nikolentzos2017multivariate}.   
(4) \texttt{Reuters21578}:  a collection of new stories from \url{nltk.corpus}.   
(5) \texttt{Polarity}: a collection of positive and negative snippets acquired from Rotten Tomatoes \cite{pang2005seeing}.  
(6) \texttt{20NewsGroups}: a collection of  news stories from \url{nltk.corpus}.   
(7) \texttt{Sixxx Requirement OBjects} (\texttt{SiROBs}): a collection of paragraphs extracted from industrial tender documents (our industrial corpus). 
See the {\it supplementary material} (Table \ref{rawtext}) for the data description and few examples texts.     
 Table \ref{datastatistics} shows the data properties and statistics. 


\subsection{Generalization}
{\bf Perplexity (PPL):} We evaluate the topic models' generative performance as a generative model  of documents 
by estimating log-probability for the test documents. 
We use the development (dev) sets of each of the seven data sets to build the corresponding models.  
We also investigate the effect of  scaling  factor ($D$) in DocNADE
\footnote{rerun: www.dmi.usherb.ca/~larocheh/code/DocNADE.zip} 
and iDocNADE models, and observe that {\it no scaling} performs better than {\it scaling}.  
See the hyperparameters for generalization in the {\it supplementary material} (Table \ref{hyperparamtersppl}), where scaling is also treated as a hyperparameter. 
A comparison is made with the {\it baseline} DocNADE and proposed iDcoNADE  using 50 or 200 topics, set by the hidden layer size $H$.   

{\bf Quantitative:} Table \ref{datastatistics}  shows the average held-out perplexity ($PPL$) per word as,\vspace{-0.3cm}  
\begin{equation*}
 PPL = \exp \big( - \frac{1}{N} \sum_{t=1}^{N} \frac{1}{|{\bf v}^t|} \log p({\bf v}^{t}) \big)  
\end{equation*}

where $N$  and $|{\bf v}^t|$ are the total number of documents and words in a document ${\bf v}^{t}$.  
The log-likelihood of the document ${\bf v}^{t}$, i.e., $\log p({\bf v}^{t})$ is obtained by 
$\mathcal{L}^{DocNADE}$ (eqn. \ref{eq:DocNADEloglikelihood}) and $\mathcal{L}^{iDocNADE}$  
(eqn. \ref{eq:iDocNADEloglikelihood}) in DocNADE and iDocNADE, respectively. 

In Table \ref{datastatistics}, we observe that DocNADE or iDocNADE performs better in 200 ($T200$) topics than $50$. 
The proposed iDocNADE achieves lower perplexity  (\underline{$601$} Vs $650$) and ({\bf $556$} Vs $596$) 
than {\it baseline} DocNADE for $50$ and $200$ topics, respectively on an average over the seven datasets.

\makeatletter
\def\labelonly{BDF}
\def\labelcheck#1{
    \edef\pgfmarshal{\noexpand\pgfutil@in@{#1}{\labelonly}}
    \pgfmarshal
    \ifpgfutil@in@[#1]\fi
}
\makeatother

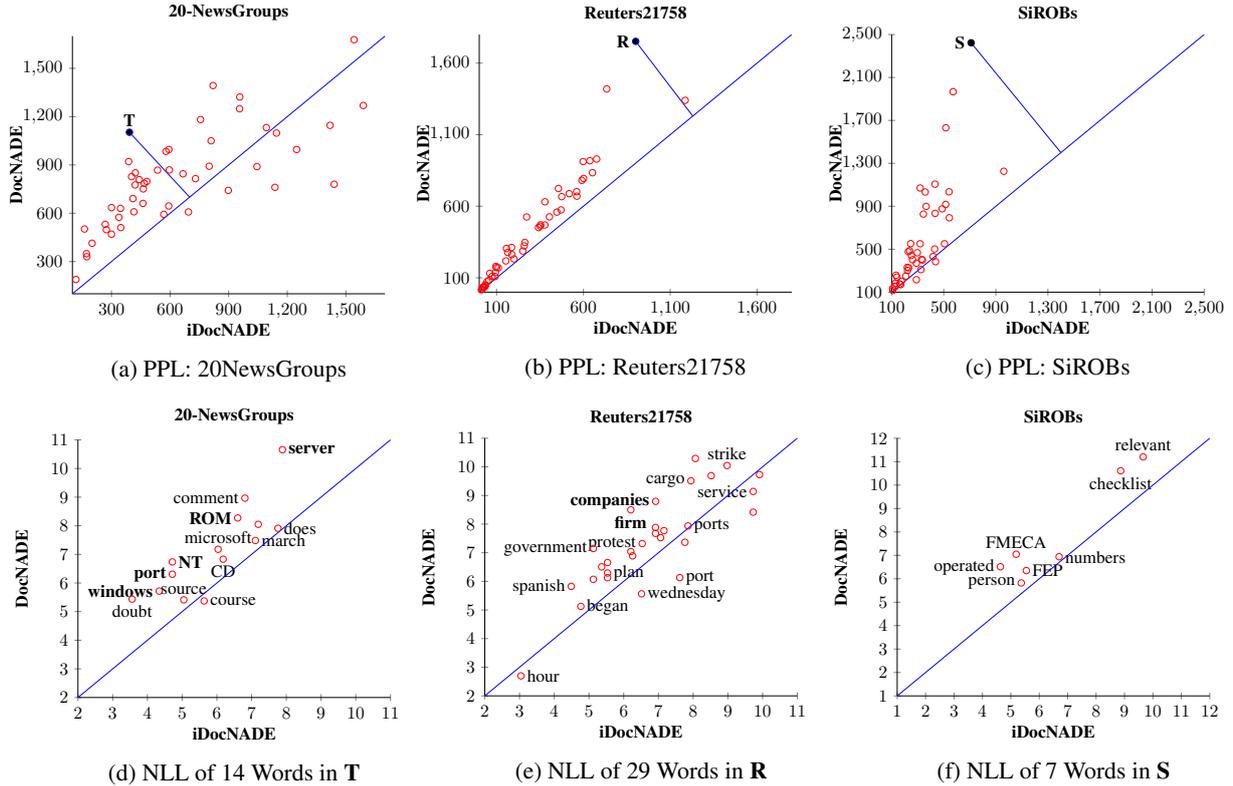
\begin{figure*}[t]
\centering
\begin{subfigure}{0.33\textwidth}
\centering
\begin{tikzpicture}[scale=0.6,trim axis left, trim axis right][baseline]
\begin{axis}[%
axis x line=bottom,
axis y line=left,
axis line style={-},
title={\bf 20-NewsGroups},
xmin=100,
xmax=1700,
ymin=100, 
ymax=1700,
/pgfplots/xtick={300,600,...,1700},
/pgfplots/ytick={300,600,...,1700},
xlabel= {\bf iDocNADE},
ylabel= {\bf DocNADE},
scatter/classes={%
    a={mark=o,draw=red}, b={draw=blue}}]
\addplot[scatter,only marks,%
    scatter src=explicit symbolic]%
table[meta=label] {
x y label
415 609 a
421 776 a
596 869 a
162  502 a
118  189 a
820 1393 a
403  827 a
1045 890 a
410 691 a
593 645 a
1542 1678 a
957 1322 a
201 414 a
346 630 a
536 868 a
481 798 a
1248 996  a
810 1050 a 
468 786 a
666 845 a
174 330 a
756  1182 a
594 996 a 
1093 1132 a
338 575 a
956 1249 a
2380 1385 a
579  984 a
1589 1269 a
274 498 a
1419 1146 a 
172  349 a
268  531 a 
463  751 a 
568  593 a 
388  921 a
1145  1098 a
300 635 a
392 1103 b
800  892 a
694 608 a
461 661 a
1440 780 a
300 470 a
348 511 a
442 809 a
422 851 a
730 815 a
1137 761 a
898 742 a
    };
\node [above] at (axis cs:  392,  1103) {\bf T};

\addplot[color=blue] coordinates {
	(100,100) (1700,1700)
};
\addplot[color=blue] coordinates {
	(392,1103) (700,700)
};
\end{axis}
\end{tikzpicture}%
\caption{PPL: 20NewsGroups} \label{PPL: 20NewsGroup}
\end{subfigure}\hspace*{\fill}%
~%
\begin{subfigure}{0.33\textwidth}
\centering
\begin{tikzpicture}[scale=0.6,trim axis left, trim axis right]
\begin{axis}[%
axis x line=bottom,
axis y line=left,
axis line style={-},
title={\bf Reuters21758},
xmin=0,
xmax=1800,
ymin=0, 
ymax=1800,
/pgfplots/xtick={100,600,...,1600},
/pgfplots/ytick={100,600,...,1600},
xlabel= {\bf iDocNADE},
ylabel= {\bf DocNADE},
scatter/classes={%
    a={mark=o,draw=red}, b={draw=blue}}]
\addplot[scatter,only marks,%
    scatter src=explicit symbolic]%
table[meta=label] {
x y label
562 670 a
518 688 a
341 451 a
28 34 a
110 170 a
561 702 a
591 781 a
475 667 a
24 33 a
354 470 a
251 287 a
263 347 a
45 74 a
901 1752 b
378 631 a
601 794 a
200 232 a
259 323 a
157 305 a
12 16 a
447 559 a
404 526 a
637 918 a
186 312 a
61 131 a
378 469 a
164 277 a
456 724 a
16 21 a
652 834 a
471 574 a
56 82 a
74 110 a
93 136 a
91 109 a
100 172 a
273 525 a
153 218 a
96 180 a
598 912 a
1186 1340 a
35 51 a
31 37 a
188 262 a
734 1419 a
19 31 a
676 930 a
36 46 a
27 39 a
351 461 a
    };
\node [left] at (axis cs:  901, 1752) {\bf R};

\addplot[color=blue] coordinates {
	(0,0) (1800,1800)
};
\addplot[color=blue] coordinates {
	(901,1752) (1230,1230)
};
\end{axis}
\end{tikzpicture}%
\caption{PPL: Reuters21758} \label{PPL: Reuters21758}
\end{subfigure}\hspace*{\fill}%
~%
~%
\begin{subfigure}{0.33\textwidth}
\centering
\begin{tikzpicture}[scale=0.6,trim axis left, trim axis right]
\begin{axis}[%
axis x line=bottom,
axis y line=left,
axis line style={-},
title={\bf SiROBs},
xmin=100,
xmax=2500,
ymin=100, 
ymax=2500,
/pgfplots/xtick={100,500,...,2500},
/pgfplots/ytick={100,500,...,2500},
xlabel= {\bf iDocNADE},
ylabel= {\bf DocNADE},
scatter/classes={%
    a={mark=o,draw=red}}]
\addplot[scatter,only marks,%
    scatter src=explicit symbolic]%
table[meta=label] {
x y label
109 116 a
162 176 a
59 34 a
128 152 a
238 486 a
319 551 a
337 399 a
290 370 a
57 91 a
572 1967 a
437 384 a
541 1035 a
134 258 a
418 432 a
141 241 a
488 876 a
98 221 a
515 917 a
261 402 a
325 310 a
434 834 a
344 827 a
365 899 a
170 170 a
516 1631 a
246 552 a
358 1033 a
97 83 a
433 1106 a
961 1226 a
220 331 a
108 134 a
129 183 a
175 202 a
430 502 a
222 302 a
297 467 a
209 250 a
506 551 a
125 156 a
228 477 a
231 327 a
543 793 a
251 441 a
89 164 a
330 406 a
96 152 a
710 2423 b
319 1070 a
290 216 a
    };
\node [left] at (axis cs:  710, 2423 ) {\bf S};
\addplot[color=blue] coordinates {
	(100,100) (2500,2500)
};
\addplot[color=blue] coordinates {
	(710,2423) (1400,1400)
};

\end{axis}
\end{tikzpicture}%
\caption{PPL: SiROBs} \label{PPL: SiROBs}
\end{subfigure}\hspace*{\fill}%
\medskip

\hspace*{\fill}%
~%
\begin{subfigure}{0.33\textwidth}
\centering
\begin{tikzpicture}[scale=0.6,trim axis right,trim axis left][baseline]
\begin{axis}[%
axis x line=bottom,
axis y line=left,
axis line style={-},
title={\bf 20-NewsGroups},
xmin=2.0,
xmax=11.0,
ymin=2.0, 
ymax=11.0,
/pgfplots/xtick={2.0,...,11.0},
/pgfplots/ytick={2.0,...,11.0},
xlabel= {\bf iDocNADE},
ylabel= {\bf DocNADE},,
scatter/classes={%
    a={mark=o,draw=red}}]
\addplot[scatter,only marks,%
    scatter src=explicit symbolic]%
table[meta=label] {
x y label
6.18075456 6.83192918 a
6.80839485 8.96412021 a
5.63295427 5.37235833 a
7.75847918 7.91195391 a
3.5600748 5.43678675 a
7.1103252 7.48475653 a
6.0352862 7.17878941 a
4.71765834 6.73556674 a
7.19305586 8.04897594 a
4.71765834 6.30657145 a
6.60034615 8.27238095 a
7.88924478 10.65654915 a
5.04651563 5.41208745 a
4.33936622 5.71744183 a
    };

\node [below] at (axis cs: 6.18075456,6.83192918) {CD};
\node [left] at (axis cs: 6.80839485, 8.96412021) {comment}; 
\node [right] at (axis cs: 5.63295427, 5.37235833) {course}; 
\node [right] at (axis cs: 7.75847918 ,7.91195391) {does}; 
\node [below] at (axis cs: 3.5600748 ,5.43678675 ) {doubt};
\node [right] at (axis cs: 7.1103252, 7.48475653 ) {march};
\node [above] at (axis cs: 6.0352862, 7.17878941) {microsoft}; 
\node [right] at (axis cs: 4.71765834, 6.73556674) {\bf NT}; 
\node [left] at (axis cs: 4.71765834, 6.30657145) {\bf port}; 
\node [left] at (axis cs: 6.60034615, 8.27238095) {\bf ROM}; 
\node [right] at (axis cs: 7.88924478, 10.6565491) {\bf server};
\node [above] at (axis cs: 5.04651563, 5.41208745) {source}; 
\node [left] at (axis cs: 4.33936622, 5.71744183) {\bf windows};

\addplot[color=blue] coordinates {
	(2.0,2.0) (11.0,11.0)
};
\end{axis}
\end{tikzpicture}%
\caption{NLL of 14 Words in \bf T} \label{NLL20newsgroup}
\end{subfigure}\hspace*{\fill}%
~%
\begin{subfigure}{0.33\textwidth}
\centering
\begin{tikzpicture}[scale=0.6,trim axis right, trim axis left]
\begin{axis}[%
axis x line=bottom,
axis y line=left,
axis line style={-},
title={\bf Reuters21758},
xmin=2.0,
xmax=11.0,
ymin=2.0, 
ymax=11.0,
/pgfplots/xtick={2.0,...,11.0},
/pgfplots/ytick={2.0,...,11.0},
xlabel= {\bf iDocNADE},
ylabel= {\bf DocNADE},
scatter/classes={%
    a={mark=o,draw=red}}]
\addplot[scatter,only marks,%
    scatter src=explicit symbolic]%
table[meta=label] {
x y label
8.06428768 10.29069646 a
9.7295979 8.41629902 a
4.76523406 5.12587523 a
7.16046003 7.7703198 a
7.93574883 9.51180985 a
6.91988634 8.79445551 a
6.20540319 8.49892026 a
5.12518561 7.14655768 a
5.36488485 6.50593951 a
3.04242776 2.69758954 a
5.53364955 6.65535857 a
6.91606603 7.88156953 a
6.20540319 7.04186241 a
9.90985577 9.72673722 a
5.53364955 6.29157032 a
7.8532203 7.93691035 a
7.61554808 6.12969682 a
6.5335862 7.3203602 a
5.53364955 6.12724089 a
9.7295979 9.1398211 a
7.76452504 7.36551685 a
8.51877906 9.69055216 a
4.48930036 5.82294644 a
7.06303388 7.52691865 a
8.9769079 10.04474953 a
6.91606603 7.67635479 a
6.25671811 6.89018505 a
6.51422361 5.5639786 a
5.12518561 6.07137121 a
    };

\node [right] at (axis cs: 4.76523406, 5.12587523 ) {began};
\node [left] at (axis cs:  7.93574883, 9.51180985 ) {cargo};
\node [left] at (axis cs:  6.91988634, 8.79445551) {\bf companies};
\node [below] at (axis cs: 6.20540319, 8.49892026) {\bf firm}; 
\node [left] at (axis cs:5.12518561, 7.14655768 ) {government};
\node [right] at (axis cs:3.04242776, 2.69758954 ) {hour};
\node [right] at (axis cs: 5.53364955, 6.29157032) {plan}; 
\node [right] at (axis cs: 7.8532203 ,7.93691035 ) {ports};
\node [right] at (axis cs: 7.61554808, 6.12969682) {port}; 
\node [left] at (axis cs: 6.5335862, 7.3203602 ) {protest};
\node [left] at (axis cs: 9.7295979, 9.1398211 ) {service};
\node [left] at (axis cs: 4.48930036, 5.82294644 ) {spanish};
\node [above] at (axis cs: 8.9769079, 10.04474953 ) {strike};
\node [right] at (axis cs: 6.51422361, 5.5639786 ) {wednesday};

\addplot[color=blue] coordinates {
	(2.0,2.0) (11.0,11.0)
};
\end{axis}
\end{tikzpicture}%
\caption{NLL of 29 Words in \bf R} \label{NLLreuters21758}
\end{subfigure}\hspace*{\fill}%
~%
~%
\begin{subfigure}{0.33\textwidth}
\centering
\begin{tikzpicture}[scale=0.6,trim axis left, trim axis right]
\begin{axis}[%
axis x line=bottom,
axis y line=left,
axis line style={-},
title={\bf SiROBs},
xmin=1.0,
xmax=12.0,
ymin=1.0, 
ymax=12.0,
/pgfplots/xtick={1.0,2.0,...,12.0},
/pgfplots/ytick={1.0, 2.0,...,12.0},
xlabel= {\bf iDocNADE},
ylabel= {\bf DocNADE},
scatter/classes={%
    a={mark=o,draw=red}}]
\addplot[scatter,only marks,%
    scatter src=explicit symbolic]%
table[meta=label] {
x y label
5.18710377306 7.04728435729 a
6.69949465199 6.9410573354 a
8.86633387532 10.6113635893 a
5.36686605535 5.82289784581 a
9.65639412069 11.2030891931 a
5.5460298076 6.34969993704 a
4.63333921335 6.51188616891 a
    };

\node [above] at (axis cs:  5.18710377306,7.04728435729) {FMECA};
\node [right] at (axis cs: 6.69949465199, 6.9410573354) {numbers};
\node [below] at (axis cs: 8.86633387532, 10.6113635893) {checklist};
\node [left] at (axis cs: 5.36686605535, 5.82289784581) {person}; 
\node [above] at (axis cs: 9.65639412069, 11.2030891931) {relevant}; 
\node [right] at (axis cs: 5.5460298076, 6.34969993704) {FEP}; 
\node [left] at (axis cs: 4.63333921335, 6.51188616891) {operated}; 

\addplot[color=blue] coordinates {
	(1.0,1.0) (12.0,12.0)
};
\end{axis}
\end{tikzpicture}%
\caption{NLL of 7 Words in \bf S} \label{NLLSiROBs}
\end{subfigure}\hspace*{\fill}%
\caption{{(a, b, c)}: PPL (200 topics) by iDocNADE and DocNADE for each of the 50 held-out documents. 
The {\it filled circle} and symbols ({\bf T}, {\bf R} and {\bf S} 
point to the document for which {\it PPL} differs by maximum,  
each for 20NewsGroups, Reuters21758 
and SiROBs datsets, respectively. 
{(d, e, f)}:  NLL of each of the words in documents marked by {\bf T}, {\bf R} and  {\bf S},  
respectively due to iDocNADE and DocNADE.} 
\label{fig:PLandIR}
\end{figure*}
\makeatletter

\makeatletter
\def\labelonly{BDF}
\def\labelcheck#1{
    \edef\pgfmarshal{\noexpand\pgfutil@in@{#1}{\labelonly}}
    \pgfmarshal
    \ifpgfutil@in@[#1]\fi
}
\makeatother

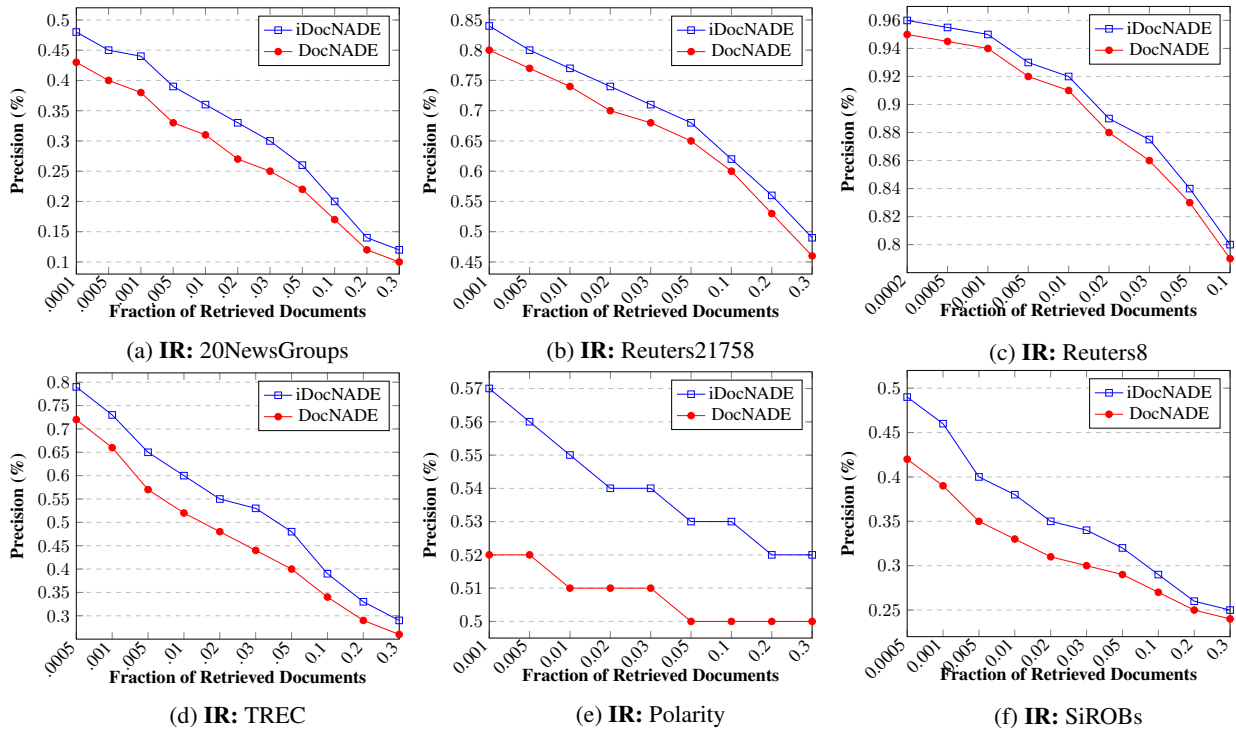
\begin{figure*}[t]
\centering
\begin{subfigure}{0.34\textwidth}
\centering
\begin{tikzpicture}[scale=0.62,trim axis left, trim axis right][baseline]
\begin{axis}[
    xlabel={\bf Fraction of Retrieved Documents},
    ylabel={\bf Precision (\%)},
    xmin=0, xmax=10,
    ymin=0.08, ymax=0.52,
    /pgfplots/ytick={.10,.15,...,.55},
    xtick={0,1,2,3,4,5,6,7,8,9,10},
    xticklabels={.0001, .0005, .001, .005, .01, .02, .03, .05, 0.1, 0.2, 0.3},
    x tick label style={rotate=48,anchor=east},
    legend pos=north east,
    ymajorgrids=true,
    grid style=dashed,
]
\addplot[
    color=blue,
    mark=square,
    ]
    plot coordinates {
    (0, 0.48)
    (1, 0.45)
    (2, 0.44)
    (3, 0.39)
    (4, 0.36)
    (5, 0.33)
    (6, 0.30)
    (7, 0.26)
    (8, 0.20)
    (9, 0.14)
    (10, 0.12)
    };

\addlegendentry{iDocNADE}

\addplot[
	color=red,
	mark=*,
	]
	plot coordinates {
    (0, 0.43)
    (1, 0.40)
    (2, 0.38)
    (3, 0.33)
    (4, 0.31)
    (5, 0.27)
    (6, 0.25)
    (7, 0.22)
    (8, 0.17)
    (9, 0.12)
    (10, 0.10)
	};
\addlegendentry{DocNADE}

\end{axis}
\end{tikzpicture}%
\caption{{\bf IR:} 20NewsGroups} \label{IR20newsgroup}
\end{subfigure}\hspace*{\fill}%
~%
\begin{subfigure}{0.33\textwidth}
\centering
\begin{tikzpicture}[scale=0.62,trim axis left, trim axis right][baseline]
\begin{axis}[
    xlabel={\bf Fraction of Retrieved Documents}, 
    ylabel={\bf Precision (\%)},
    xmin=0, xmax=8,
    ymin=0.43, ymax=0.87,
   /pgfplots/ytick={.45,.50,...,.90},
    xtick={0,1,2,3,4,5,6,7,8},
    xticklabels={0.001,0.005, 0.01, 0.02, 0.03, 0.05, 0.1, 0.2, 0.3},
    x tick label style={rotate=45,anchor=east},
    legend pos=north east,
    ymajorgrids=true,
    grid style=dashed,
]
\addplot[
    color=blue,
    mark=square,
    ]
    plot coordinates {
    (0, 0.84)
    (1, 0.80)
    (2, 0.77)
    (3, 0.74)
    (4, 0.71)
    (5, 0.68)
    (6, 0.62)
    (7, 0.56)
    (8, 0.49)
    };

\addlegendentry{iDocNADE}

\addplot[
	color=red,
	mark=*,
	]
	plot coordinates {
    (0, 0.80)
    (1, 0.77)
    (2, 0.74)
    (3, 0.70)
    (4, 0.68)
    (5, 0.65)
    (6, 0.60)
    (7, 0.53)
    (8, 0.46)
	};
\addlegendentry{DocNADE}

\end{axis}
\end{tikzpicture}%
\caption{{\bf IR:} Reuters21758} \label{IRReuters21758}
\end{subfigure}\hspace*{\fill}%
~%
\hspace*{\fill}
~%
\begin{subfigure}{0.33\textwidth}
\centering
\begin{tikzpicture}[scale=0.62,trim axis left, trim axis right][baseline]
\begin{axis}[
    xlabel={\bf Fraction of Retrieved Documents},
    ylabel={\bf Precision (\%)},
    xmin=0, xmax=8,
    ymin=0.78, ymax=0.97,
   /pgfplots/ytick={.80,.82,...,1.0},
    xtick={0,1,2,3,4,5,6,7,8},
    xticklabels={0.0002, 0.0005,0.001,0.005, 0.01, 0.02, 0.03, 0.05, 0.1},
    x tick label style={rotate=45,anchor=east},
    legend pos=north east,
    ymajorgrids=true,
    grid style=dashed,
]
\addplot[
    color=blue,
    mark=square,
    ]
    plot coordinates {
    (0, 0.96)
    (1, 0.955)
    (2, 0.95)
    (3, 0.93)
    (4, 0.92)
    (5, 0.89)
    (6, 0.875)
    (7, 0.84)
    (8, 0.80)
    };

\addlegendentry{iDocNADE}

\addplot[
	color=red,
	mark=*,
	]
	plot coordinates {
    (0, 0.95)
    (1, 0.945)
    (2, 0.94)
    (3, 0.92)
    (4, 0.91)
    (5, 0.88)
    (6, 0.86)
    (7, 0.83)
    (8, 0.79)
	};
\addlegendentry{DocNADE}

\end{axis}
\end{tikzpicture}%
\caption{{\bf IR:} Reuters8} \label{IRReuters8}
\end{subfigure}\hspace*{\fill}%
~%

\begin{subfigure}{0.34\textwidth}
\centering
\begin{tikzpicture}[scale=0.62,trim axis left, trim axis right][baseline]
\begin{axis}[
    xlabel={\bf Fraction of Retrieved Documents},
    ylabel={\bf Precision (\%)},
    xmin=0, xmax=9,
    ymin=0.25, ymax=0.82,
    /pgfplots/ytick={.30,.35,...,.85},
    xtick={0,1,2,3,4,5,6,7,8,9},
    xticklabels={.0005, .001, .005, .01, .02, .03, .05, 0.1, 0.2, 0.3},
    x tick label style={rotate=48,anchor=east},
    legend pos=north east,
    ymajorgrids=true,
    grid style=dashed,
]
\addplot[
    color=blue,
    mark=square,
    ]
    plot coordinates {
    (0, 0.79)
    (1, 0.73)
    (2, 0.65)
    (3, 0.60)
    (4, 0.55)
    (5, 0.53)
    (6, 0.48)
    (7, 0.39)
    (8, 0.33)
    (9, 0.29)
    };

\addlegendentry{iDocNADE}

\addplot[
	color=red,
	mark=*,
	]
	plot coordinates {
    (0, 0.72)
    (1, 0.66)
    (2, 0.57)
    (3, 0.52)
    (4, 0.48)
    (5, 0.44)
    (6, 0.40)
    (7, 0.34)
    (8, 0.29)
    (9, 0.26)
	};
\addlegendentry{DocNADE}

\end{axis}
\end{tikzpicture}%
\caption{{\bf IR:} TREC} \label{IRTREC}
\end{subfigure}\hspace*{\fill}%
~%
\begin{subfigure}{0.33\textwidth}
\centering
\begin{tikzpicture}[scale=0.62,trim axis left, trim axis right][baseline]
\begin{axis}[
    xlabel={\bf Fraction of Retrieved Documents},
    ylabel={\bf Precision (\%)},
    xmin=0, xmax=8,
    ymin=0.495, ymax=0.575,
   /pgfplots/ytick={.50,.51,...,.60},
    xtick={0,1,2,3,4,5,6,7,8},
    xticklabels={0.001,0.005, 0.01, 0.02, 0.03, 0.05, 0.1, 0.2, 0.3},
    x tick label style={rotate=45,anchor=east},
    legend pos=north east,
    ymajorgrids=true,
    grid style=dashed,
]
\addplot[
    color=blue,
    mark=square,
    ]
    plot coordinates {
    (0, 0.57)
    (1, 0.56)
    (2, 0.55)
    (3, 0.54)
    (4, 0.54)
    (5, 0.53)
    (6, 0.53)
    (7, 0.52)
    (8, 0.52)
    };

\addlegendentry{iDocNADE}

\addplot[
	color=red,
	mark=*,
	]
	plot coordinates {
    (0, 0.52)
    (1, 0.52)
    (2, 0.51)
    (3, 0.51)
    (4, 0.51)
    (5, 0.50)
    (6, 0.50)
    (7, 0.50)
    (8, 0.50)
	};
\addlegendentry{DocNADE}

\end{axis}
\end{tikzpicture}%
\caption{{\bf IR:} Polarity} \label{IRPolarity}
\end{subfigure}\hspace*{\fill}%
~%
\hspace*{\fill}
~%
\begin{subfigure}{0.33\textwidth}
\centering
\begin{tikzpicture}[scale=0.62,trim axis left, trim axis right][baseline]
\begin{axis}[
    xlabel={\bf Fraction of Retrieved Documents},
    ylabel={\bf Precision (\%)},
    xmin=0, xmax=9,
    ymin=0.22, ymax=0.52,
   /pgfplots/ytick={.25,.30,...,.55},
    xtick={0,1,2,3,4,5,6,7,8,9},
    xticklabels={0.0005, 0.001, 0.005, 0.01, 0.02, 0.03, 0.05, 0.1, 0.2, 0.3},
    x tick label style={rotate=45,anchor=east},
    legend pos=north east,
    ymajorgrids=true,
    grid style=dashed,
]
\addplot[
    color=blue,
    mark=square,
    ]
    plot coordinates {
    (0, 0.49)
    (1, 0.46)
    (2, 0.40)
    (3, 0.38)
    (4, 0.35)
    (5, 0.34)
    (6, 0.32)
    (7, 0.29)
    (8, 0.26)
    (9, 0.25)
    };

\addlegendentry{iDocNADE}

\addplot[
	color=red,
	mark=*,
	]
	plot coordinates {
    (0, 0.42)
    (1, 0.39)
    (2, 0.35)
    (3, 0.33)
    (4, 0.31)
    (5, 0.30)
    (6, 0.29)
    (7, 0.27)
    (8, 0.25)
    (9, 0.24)
	};
\addlegendentry{DocNADE}

\end{axis}
\end{tikzpicture}%
\caption{{\bf IR:} SiROBs} \label{IRSiROBs}
\end{subfigure}\hspace*{\fill}%
~%
\caption{DocNADE Vs iDocNADE: Document retrieval performance (precision) on 
six data sets at different retrieval fractions. Observe different y-axis scales.}
\label{fig:docretrieval}
\end{figure*}
\makeatletter

{\bf Inspection:} We quantify the use of context information in learning informed document representations. 
For the three datasets (namely 20-NewsGroups, Reuters21758 and SiROBs), 
we randomly select 50 held-out documents from their test sets and compare 
(Figure \ref{PPL: 20NewsGroup}, \ref{PPL: Reuters21758} and \ref{PPL: SiROBs}) the {\it PPL}  
for each of the held-out documents under the learned (optimal) 200-dimensional DocNADE and iDocNADE.   
Observe that iDocNADE achieves lower {\it PPL} for the majority of the documents.  
The {\it filled} circle(s) 
points to the document for which {\it PPL} differs by a maximum between iDocNADE and DocNADE.  
For each dataset, we select the corresponding document and compute the negative log-likelihood ({\it NLL}) for every word.  
Figure \ref{NLL20newsgroup}, \ref{NLLreuters21758} and \ref{NLLSiROBs}  
show that the {\it NLL}  for the majority of the words is lower (better) in iDocNADE than DocNADE. 

\begin{table}[t]
\centering
\small
\renewcommand*{\arraystretch}{1.2}
\resizebox{0.49\textwidth}{!}{
\begin{tabular}{c|c|c|c}
{\bf model} &  {\bf 20NewsGroups} & {\bf  Reuters21758}  & {\bf  SiROBs} \\ \hline
{\it DocNADE} &              0.705       &        0.573                          &     0.400             \\
{\it iDocNADE} &          \underline{0.710}           &        \underline{0.581}                      &         \underline{0.409}         
\end{tabular}}
\caption{Average coherence over all the topics (200) learned by DocNADE and iDocNADE}
\label{coherencescores}
\end{table}
\begin{table}[t]
\centering
\renewcommand*{\arraystretch}{1.15}
\resizebox{0.49\textwidth}{!}{
\begin{tabular}{cc|cc}
\multicolumn{2}{c|}{\bf 20NewsGroups} & \multicolumn{2}{c}{\bf Reuters21758} \\
{\it DocNADE}        & {\it iDocNADE}        & {\it DocNADE}         & {\it iDocNADE}         \\ \hline
  jesus, bible             &        jesus, jews                     &        bank, dollar          &         billion, record        \\
  christian, christians             &      christians, christ           &      billion, deficit            &        credit, dollar         \\
  religion, book             &    religion, bible                  &      rate, dealer            &          fall, loss       \\
   government, faith            &    high, religious              &      trade, loan            &       trade, export          \\
  science, jewish             &   church, christian                   &     loss, money             &         sale, prime        \\ \hline 
      0.700         &        \underline{0.707}                  &          0.575        &     \underline{0.628}            
\end{tabular}}
\caption{Topics (top 10 words) with coherence}
\label{topiccoherence}
\end{table}

\subsection{Interpretability: Topic Coherence}
Beyond PPL, 
we compute topic coherence 
\cite{Chang:82,Newman:82,Das:82,Gupta:85} 
to assess the meaningfulness of the 
underlying topics captured. 
We choose the coherence measure proposed by \newcite{Michael:82}   
that identifies context features for each topic word using a sliding window over the reference corpus. The higher scores imply more coherent topics.

{\bf Quantitative:} We use gensim module\footnote{radimrehurek.com/gensim/models/coherencemodel.html} 
({\it coherence type} = $c\_v$) to estimate coherence 
for each of the 200 topics (top-10 words), captured by the least perplexed DocNADE and iDocNADE. 
Table \ref{coherencescores} shows average coherence over 200 topics, 
where the high scores suggest more coherent topics in iDocNADE compared to DocNADE.

{\bf Qualitative:} Table \ref{topiccoherence} illustrates example topics  with coherence by DocNADE and iDocNADE. 


\subsection{Applicability: Document Retrieval}
To evaluate the quality of the learned representations, 
we perform a document retrieval task using the six datasets and their label information.
We use the experimental setup similar to  \newcite{Hugo:82}, 
where all test documents are treated as queries to retrieve a fraction of the closest documents in
the original training set using cosine similarity measure between their representations (eqn. \ref{eq:documentrepiDocNADE} in iDocNADE 
and $\overrightarrow{\bf h}_D$ in DocNADE).  
To compute retrieval precision for each fraction (e.g., $0.0001$, $0.005$, $0.01$, $0.02$, $0.05$, $0.1$, $0.2$, $0.5$, etc.), 
we average the number of retrieved training documents with the same label as the query. 
For multi-label data (Reuters21758 and SiROBs), we average the precision scores over multiple labels for each query.  
Since \newcite{Salakhutdinov:82} 
and  \newcite{Hugo:82} 
showed that RSM and DocNADE strictly outperforms 
LDA on this task, we only compare DocNADE and iDocNADE.  
Figures \ref{IR20newsgroup},  \ref{IRReuters21758},  \ref{IRReuters8}, \ref{IRTREC}, \ref{IRPolarity}, \ref{IRSiROBs} show 
the average precision for the retrieval task on 20NewsGroups, Reuters21758, Reuters8,  TREC, Polarity and SiROBs datasets, respectively. 
Observe that iDocNADE outperforms DocNADE in precision at different retrieval fractions (particularly for the top retrievals) for all the single and multi-labeled datasets from different domains.   
For instance, at retrieval fraction $0.02$, we (in Table \ref{datastatistics}) report a gain of $9.6$\% ($.57$ Vs $.52$) in precision on an average over the six datasets, compared to DocNADE. 

\begin{table}[t]
\centering
\renewcommand*{\arraystretch}{1.2}
\resizebox{0.49\textwidth}{!}{
\begin{tabular}{c||cc|cc|cc}
\hline
\multirow{2}{*}{\bf Data}   & \multicolumn{2}{c|}{\bf DocNADE}      & \multicolumn{2}{c|}{\bf iDocNADE}    & \multicolumn{2}{c}{\bf Word2vecDoc}\\ \cline{2-7}
                             & {$ F1$}       & {$Acc$}  & {$ F1$}    & {$Acc$}  & {$ F1$}    & {$ Acc$}   \\ \hline
{\it  TREC}    &     .780                  &  .778                 &      .797        &      .802                   &   {\bf  .825}   &      {\bf .822}     \\   
{\it  Reuters8}    &     .899                  &  .962                 &      {\bf .920}        &      {\bf .970}              &       .910  &     .966      \\   
{\it  Reuters21758}    &     .329                  &  .644                 &      .310        &      .637          &    {\bf .402}  &     {\bf  .769}           \\ 
{\it  Polarity}    &     .513                  &  .513                 &      .698        &      .698              &      {\bf .725 }     &    {\bf .725}      \\   
{\it  20NewsGroups}    &     .503                  &  .523                 &      {\bf .521}        &      {\bf .537}              &      .512      &    .528      \\   
{\it  SiROBs}    &     .233                  &  .327                 &      .245        &      .335              &      .296      &    .384      \\   \hline
{\it  Average}    & .543                  &  .625                 &     \underline{.582}        &      \underline{.663}              &      .611      &    .699   
\end{tabular}}
\caption{Supervised classification using word representations (${\bf W}$) learned in DocNADE and iDocNADE.  
Word2vecDoc: sum of embedding vector from pretrained Word2vec for each word in the document. $Acc$: Accuracy. $F1$ is macro-averaged.}
\label{state-of-the-art-classification}
\end{table}

To perform document retrieval, we use the same train/development/test split of documents discussed in Table \ref{datastatistics} for 
all the datasets during learning. 
For model selection, we use the development set as the query set and use the average precision at 0.02\% retrieved documents as the
performance measure. We train DocNADE and iDocNADE models with $200$ topics and perform stochastic gradient descent for 2000 training passes with different learning rates. Note that the labels are not used during training. The class labels are only used to check if the retrieved documents have the same class label as the query document.  
See Table \ref{hyperparamtersIR} for the hyperparameters in the document retrieval task. 


\begin{table*}[t]
\centering
\renewcommand*{\arraystretch}{1.15}
\resizebox{0.99\textwidth}{!}{
\begin{tabular}{ccc|ccc|ccc|ccc|ccc}
\hline
\multicolumn{15}{c}{\bf Data: 20NewsGroups} \\ \hline
\multicolumn{3}{c|}{\bf book}   & \multicolumn{3}{c|}{\bf jesus}   & \multicolumn{3}{c|}{\bf windows}   & \multicolumn{3}{c|}{\bf gun}   & \multicolumn{3}{c}{\bf religion} \\ 
{\it neighbors} & {$s_i$}  & {$s_g$} & {\it neighbors} & {$s_i$} & {$s_g$} & {\it neighbors} & {$s_i$}    & {$s_g$} & {\it neighbors} & {$s_i$}  & {$s_g$} & {\it neighbors} & {$s_i$}   & {$s_g$}\\  \hline
       books   &  .61    &    .74    &  christ      &   .86       &   .64       &  dos       &  .74         &    .28          &  guns   &   .72    &   .78     & religious    &   .75    &   .72\\
       reference   & .52    &  .18      &   god     &   .78       &  .52        &  files         &   .63         &  .23            &  firearms   &  .63       &  .73   &  christianity   &   .71     &   .53\\
       published   &    .46  &  .39      & christians      & .74         &    .58      &  version         &   .59        &    .10          &  criminal      &  .63   &  .22   &   beliefs  &   .66      &   .49\\
       reading   &    .45   &    .38    & faith      &     .71       &   .20           & file         &  .59      &   .15     &  crime   &    .62     &   .28  &   christian  &   .68      &   .45\\
       author  &   .44  &   .54     & bible      &   .71      &   .35       &    unix     &  .52          &   .38           &  police   & .61      &   .34  &   religions  &   .67      &   .75 \\ \hline \hline
\multicolumn{15}{c}{\bf Data: Reuters21758} \\  \hline
\multicolumn{3}{c|}{\bf buy}   & \multicolumn{3}{c|}{\bf meat}   & \multicolumn{3}{c|}{\bf gold}   & \multicolumn{3}{c|}{\bf rise}   & \multicolumn{3}{c}{\bf cotton} \\ 
{\it neighbors} & {$s_i$}  & {$s_g$} & {\it neighbors} & {$s_i$} & {$s_g$} & {\it neighbors} & {$s_i$}  & {$s_g$} & {\it neighbors} & {$s_i$}  & {$s_g$} & {\it neighbors} & {$s_i$} & {$s_g$}\\  \hline
  deal        &  .70    &  .27     &   pork     &   .63       &  .72         &   mine      &  .79       &  .43            &   fall  &   .83   &  .55       & sorghum    &   .71   &  .55 \\
  sign        &  .65    &  .23     &  beef      &   .57       &  .73         &    ore     &  .79       &  .47             &   rose  &   .82   &  .59       &  crop   &   .67   &  .47 \\
  bought        &  .63    &  .71      &   ban     &   .50       &  .12        &   silver      &  .79       &  .83             &  fell   &   .75   &  .40     & usda    &   .66   &  .22   \\
  own        &  .58    &  .28      &   livestock     &   .49       &  .47         &    assay     &  .76       &  .30           & drop    &   .70   &  .63       & soybean    &   .66   &  .56  \\
  sell        &  .58    &  .83      &  meal      &   .47       &  .43          &     feet    &  .72       &  .08           &  growth   &   .66  &  .44      &  bale   &   .65   &  .47  \\ \hline
\end{tabular}}
\caption{The five nearest neighbors in the word representation space learned by iDocNADE for 20NewsGroups and Reuters21758 datasets.  
$s_i$: Cosine similarity between the word vectors from iDocNADE, for instance vectors of {\it jesus} and {\it god}. 
$s_g$: Cosine Similarity in embedding vectors from word2vec . 
See the full list of neighbors for each word of the vocabulary in the {\it supplementary material} for both the datasets.}
\label{neighbors}
\end{table*}

\subsection{Applicability: Document Categorization}
Beyond the document retrieval, we perform text categorization to  measure the quality of word vectors learned in the topic models. 
We consider the same experimental setup as in the document retrieval task and extract the embedding matrix ${\bf W} \in \mathbb{R}^{H \times K}$ learned in DocNADE and iDocNADE during retrieval training, where $H$ (=200) is the hidden dimension and represents an embedding vector for each word in the vocabulary of size $K$ (Table \ref{datastatistics}).  
For each of the six datasets with label information, we compute a document representation by summing \cite{joulin2016bag} its word vectors,   
obtained as the columns ${W}_{:, v_i}$  for each word $v_i$. To perform document categorization, we employ 
a logistic regression classifier\footnote{\url{scikit-learn.org/}} 
with $L2$ regularization, parameterized by [0.01, 0.1, 1.0, 10.0]. 
We use the development set to find the optimal regularization parameter. 
Table \ref{state-of-the-art-classification} show that iDocNADE achieves higher $F1$ and classification accuracy over DocNADE.    
We show a gain of  $7.2$\% ($.582$ Vs $.543$) in $F1$ on an average over the six datasets. 

We also quantify the quality of word representations learned in iDocNADE only using the corpus documents. To do so, we compute document 
representations by summing the pre-trained word vectors from word2vec\footnote{\url{code.google.com/archive/p/word2vec/}} \cite{Mikolov:82} and 
perform classification (Word2vecDoc).  
Table \ref{state-of-the-art-classification} shows that iDocNADE achieves higher scores  than Word2vecDoc  for classification
on two datasets (20NewsGroups and Reuters8), suggesting it's competence in learning meaningful representations even in smaller corpus.  

\subsection{Inspection of Learned Representations}
To analyze the meaningful semantics captured, we perform a qualitative inspection of the learned representations by the topic models.   
Table \ref{topiccoherence} shows topics for 20NewsGroups and Reuters21758 that could be interpreted as {\it religion} and {\it trading}, 
which are (sub)categories in the data, confirming that meaningful topics are captured. 

For word level inspection, 
we extract {\it word representations} using the columns $W_{:,v_i}$ as the vector (200 dimension) representation 
of each word $v_i$, learned by iDocNADE using 20NewsGroups and Reuters21758 datasets.  
Figure \ref{neighbors} shows the five nearest neighbors of some selected words in this space 
and their corresponding  similarity scores. 
We also compare similarity in word vectors from iDocNADE and pre-trained word2vec 
embeddings (see demo: \url{bionlp-www.utu.fi/wv_demo/}), 
again confirming that meaningful word representations are learned. 

See the {\it  supplementary material} for the top-20 neighbors of each in the vocabulary, extracted by iDocNADE using 20NewsGroups and 
Reusters21758 datasets.

\begin{table}[t]
\centering
\small
\renewcommand*{\arraystretch}{1.25}
\resizebox{0.49\textwidth}{!}{
\begin{tabular}{r|cc|cc}
\multirow{1}{*}{\bf Pre-trained}    & \multicolumn{2}{c|}{\bf  In-Domain}      & \multicolumn{2}{c}{\bf Out-of-Domain}  \\
\multirow{1}{*}{\bf Model}    & \multicolumn{2}{c|}{\it  Eval: 20NewsGroups}         & \multicolumn{2}{c}{\it  Eval: SiROBs}  \\ \cline{2-5}
\multirow{1}{*}{\bf on Data}      & {\it DocNADE}      & {\it iDocNADE}           & {\it DocNADE}      & {\it iDocNADE} \\ \hline
\multirow{1}{*}{\it  Reuters21758}       &       6480                   &            \underline{5627}             &        7312                &  	\underline{7075}                                                                                                     
\end{tabular}}
\caption{Transfer Learning performance (PPL) of DocNADE and iDocNADE for {\it 20NewsGroups} and {\it SiROBs} data sets. Eval: Evaluation
}
\label{Transferlearning}
\end{table}

\subsection{Transfer Learning Generalization}
We train DocNADE and iDocNADE on  Reuters21758 and evaluate both models on 20NewsGroups and SiROBs test sets, to assess {\it in}- and {\it out-of-domain} transfer learning capabilities. 
Table \ref{Transferlearning} shows that iDocNADE obtains lower perplexity than DocNADE, suggesting a better generalization.   

{\it In-Domain}: Trained models from Reuters21758  data and evaluate on 20NewsGroup test data from the same news domain. 
{\it Out-of-Domain}: Trained models from Reuters21758 data and evaluate on  SiROBs test data from industrial domain. 
 
\section{Conclusion}
We have shown that leveraging contextual information in our proposed topic model {\it iDocNADE} results in learning 
better document and word representations, and improves generalization, 
interpretability of topics and its applicability in document retrieval and classification. 

\section*{Acknowledgments}
We express appreciation for our colleagues Ulli Waltinger, Bernt Andrassy, Mark Buckley,  
Stefan Langer, Thorsten Fuehring, Subbu Rajaram, Yatin Chaudhary and Khushbu Saxena  
for their in-depth review comments. 
This research was supported by Bundeswirtschaftsministerium ({\tt bmwi.de}), grant 01MD15010A (Smart Data Web) 
at Siemens AG- CT Machine Intelligence, Munich Germany. 

\bibliographystyle{emnlp2018}
\bibliography{emnlp2018}

\begin{thebibliography}{26}
\expandafter\ifx\csname natexlab\endcsname\relax\def\natexlab#1{#1}\fi

\bibitem[{Bengio et~al.(2003)Bengio, Ducharme, Vincent, and Jauvin}]{Bengio:82}
Yoshua Bengio, Rejean Ducharme, Pascal Vincent, and Christian Jauvin. 2003.
\newblock A neural probabilistic language model.
\newblock In \emph{Journal of Machine Learning Research 3}, pages
  1137–--1155.

\bibitem[{Blei et~al.(2003)Blei, Ng, and Jordan}]{Blei:81}
D.~Blei, A.~Ng, and M.~Jordan. 2003.
\newblock Latent dirichlet allocation.
\newblock pages 993--1022.

\bibitem[{Chang et~al.(2009)Chang, Boyd-Graber, Wang, Gerrish, and
  Blei.}]{Chang:82}
Jonathan Chang, Jordan Boyd-Graber, Chong Wang, Sean Gerrish, and David~M.
  Blei. 2009.
\newblock Reading tea leaves: How humans interpret topic models.
\newblock In \emph{In Neural Information Processing Systems (NIPS)}.

\bibitem[{Das et~al.(2015)Das, Zaheer, and Dyer}]{Das:82}
Rajarshi Das, Manzil Zaheer, and Chris Dyer. 2015.
\newblock Gaussian lda for topic models with word embeddings.
\newblock In \emph{Proceedings of the 53rd Annual Meeting of the Association
  for Computational Linguistics and the 7th International Joint Conference on
  Natural Language Processing}. Association for Computational Linguistics.

\bibitem[{Elman(1990)}]{elman1990finding}
Jeffrey~L Elman. 1990.
\newblock Finding structure in time.
\newblock \emph{Cognitive science}, 14(2):179--211.

\bibitem[{Gupta et~al.(2018)Gupta, Rajaram, Sch{\"u}tze, and
  Andrassy}]{Gupta:85}
Pankaj Gupta, Subburam Rajaram, Hinrich Sch{\"u}tze, and Bernt Andrassy. 2018.
\newblock Deep temporal-recurrent-replicated-softmax for topical trends over
  time.
\newblock In \emph{Proceedings of the 2018 Conference of the North American
  Chapter of the Association for Computational Linguistics: Human Language
  Technologies, Volume 1 (Long Papers)}, volume~1, pages 1079--1089, New
  Orleans, USA. Association of Computational Linguistics.

\bibitem[{Gupta et~al.(2015{\natexlab{a}})Gupta, Runkler, Adel, Andrassy,
  Zimmermann, and Sch{\"u}tze}]{Gupta:89}
Pankaj Gupta, Thomas Runkler, Heike Adel, Bernt Andrassy, Hans-Georg
  Zimmermann, and Hinrich Sch{\"u}tze. 2015{\natexlab{a}}.
\newblock Deep learning methods for the extraction of relations in natural
  language text.
\newblock Technical report, Technical University of Munich, Germany.

\bibitem[{Gupta et~al.(2015{\natexlab{b}})Gupta, Runkler, and
  Andrassy}]{Gupta:86}
Pankaj Gupta, Thomas Runkler, and Bernt Andrassy. 2015{\natexlab{b}}.
\newblock Keyword learning for classifying requirements in tender documents.
\newblock Technical report, Technical University of Munich, Germany.

\bibitem[{Gupta et~al.(2016)Gupta, Sch{\"u}tze, and Andrassy}]{Gupta:82}
Pankaj Gupta, Hinrich Sch{\"u}tze, and Bernt Andrassy. 2016.
\newblock Table filling multi-task recurrent neural network for joint entity
  and relation extraction.
\newblock In \emph{Proceedings of COLING 2016, the 26th International
  Conference on Computational Linguistics: Technical Papers}, pages
  2537–--2547.

\bibitem[{Gupta et~al.(2015{\natexlab{c}})Gupta, Sivalingam, P{\"o}lsterl, and
  Navab}]{Gupta:91}
Pankaj Gupta, Udhayaraj Sivalingam, Sebastian P{\"o}lsterl, and Nassir Navab.
  2015{\natexlab{c}}.
\newblock Identifying patients with diabetes using discriminative restricted
  boltzmann machines.
\newblock Technical report, Technical University of Munich, Germany.

\bibitem[{Hinton(2002)}]{Hinton:83}
Geoffrey~E. Hinton. 2002.
\newblock Training products of experts by minimizing contrastive divergence.
\newblock In \emph{Neural Computation}, pages 1771--1800.

\bibitem[{Joulin et~al.(2016)Joulin, Grave, Bojanowski, and
  Mikolov}]{joulin2016bag}
Armand Joulin, Edouard Grave, Piotr Bojanowski, and Tomas Mikolov. 2016.
\newblock Bag of tricks for efficient text classification.
\newblock \emph{arXiv preprint arXiv:1607.01759}.

\bibitem[{Larochelle and Lauly(2012)}]{Hugo:82}
Hugo Larochelle and Stanislas Lauly. 2012.
\newblock A neural autoregressive topic model.
\newblock In \emph{Proceedings of the Advances in Neural Information Processing
  Systems 25 (NIPS 2012)}. NIPS.

\bibitem[{Larochelle and Murray(2011)}]{Hugo:84}
Hugo Larochelle and Ian Murray. 2011.
\newblock The neural autoregressive distribution estimato.
\newblock In \emph{Proceedings of the 14th International Conference on
  Artificial Intelligence and Statistics (AISTATS 2011)}, pages 29--–37.
  JMLR.

\bibitem[{Li and Roth(2002)}]{li2002learning}
Xin Li and Dan Roth. 2002.
\newblock Learning question classifiers.
\newblock In \emph{Proceedings of the 19th international conference on
  Computational linguistics-Volume 1}, pages 1--7. Association for
  Computational Linguistics.

\bibitem[{Manning and Sch{\"u}tze(1999)}]{Manning:82}
Christopher~D Manning and Hinrich Sch{\"u}tze. 1999.
\newblock Foundations of statistical natural language processing.
\newblock Cambridge MA: The MIT Press.

\bibitem[{Mikolov et~al.(2013)Mikolov, Chen, Corrado, and Dean}]{Mikolov:82}
Tomas Mikolov, Kai Chen, Greg Corrado, and Jeffrey Dean. 2013.
\newblock Efficient estimation of word representations in vector space.
\newblock In \emph{In Proceedings of Workshop at ICLR,}.

\bibitem[{Mousa and Schuller(2017)}]{Mousa:82}
Amr El-Desoky Mousa and Bj{\"o}rn Schuller. 2017.
\newblock Contextual bidirectional long short-term memory recurrent neural
  network language models: A generative approach to sentiment analysis.
\newblock In \emph{Proceedings of the 15th Conference of the European Chapter
  of the Association for Computational Linguistics}, pages 1023–--1032.
  Association for Computational Linguistics.

\bibitem[{Newman et~al.(2009)Newman, Karimi, and Cavedon}]{Newman:82}
David Newman, Sarvnaz Karimi, and Lawrence Cavedon. 2009.
\newblock External evaluation of topic models.
\newblock In \emph{Proceedings of the 14th Australasian Document Computing
  Symposium}.

\bibitem[{Nikolentzos et~al.(2017)Nikolentzos, Meladianos, Rousseau, Stavrakas,
  and Vazirgiannis}]{nikolentzos2017multivariate}
Giannis Nikolentzos, Polykarpos Meladianos, Fran{\c{c}}ois Rousseau, Yannis
  Stavrakas, and Michalis Vazirgiannis. 2017.
\newblock Multivariate gaussian document representation from word embeddings
  for text categorization.
\newblock In \emph{Proceedings of the 15th Conference of the European Chapter
  of the Association for Computational Linguistics: Volume 2, Short Papers},
  volume~2, pages 450--455.

\bibitem[{Pang and Lee(2005)}]{pang2005seeing}
Bo~Pang and Lillian Lee. 2005.
\newblock Seeing stars: Exploiting class relationships for sentiment
  categorization with respect to rating scales.
\newblock In \emph{Proceedings of the 43rd annual meeting on association for
  computational linguistics}, pages 115--124. Association for Computational
  Linguistics.

\bibitem[{R{\"o}der et~al.(2015)R{\"o}der, Both, and Hinneburg}]{Michael:82}
Michael R{\"o}der, Andreas Both, and Alexander Hinneburg. 2015.
\newblock Exploring the space of topic coherence measures.
\newblock In \emph{Proceedings of the WSDM}. ACM.

\bibitem[{Salakhutdinov and Hinton(2009)}]{Salakhutdinov:82}
Ruslan Salakhutdinov and Geoffrey Hinton. 2009.
\newblock Replicated softmax: an undirected topic model.
\newblock In \emph{Proceedings of the Advances in Neural Information Processing
  Systems 22 (NIPS 2009)}, pages 1607--1614. NIPS.

\bibitem[{Vu et~al.(2016{\natexlab{a}})Vu, Adel, Gupta, and
  Sch{\"u}tze}]{Gupta:87}
Ngoc~Thang Vu, Heike Adel, Pankaj Gupta, and Hinrich Sch{\"u}tze.
  2016{\natexlab{a}}.
\newblock Combining recurrent and convolutional neural networks for relation
  classification.
\newblock In \emph{Proceedings of the North American Chapter of the Association
  for Computational Linguistics: Human Language Technologies}, pages 534--539,
  San Diego, California USA. Association for Computational Linguistics.

\bibitem[{Vu et~al.(2016{\natexlab{b}})Vu, Gupta, Adel, and
  Sch{\"u}tze}]{Vu:82}
Ngoc~Thang Vu, Pankaj Gupta, Heike Adel, and Hinrich Sch{\"u}tze.
  2016{\natexlab{b}}.
\newblock Bi-directional recurrent neural network with ranking loss for spoken
  language understanding.
\newblock In \emph{Proceedings of IEEE/ACM Trans. on Audio, Speech, and
  Language Processing (ICASSP)}. IEEE.

\bibitem[{Zheng et~al.(2016)Zheng, Zhang, and Larochelle}]{Hugo:83}
Yin Zheng, Yu-Jin Zhang, and Hugo Larochelle. 2016.
\newblock A deep and autoregressive approach for topic modeling of multimodal
  data.
\newblock In \emph{IEEE transactions on pattern analysis and machine
  intelligence}, pages 1056--1069. IEEE.

\end{thebibliography}


\begin{thebibliography}{}

\bibitem[\protect\citename{Aho and Ullman}1972]{Aho:72}
Alfred~V. Aho and Jeffrey~D. Ullman.
\newblock 1972.
\newblock {\em The Theory of Parsing, Translation and Compiling}, volume~1.
\newblock Prentice-{Hall}, Englewood Cliffs, NJ.

\bibitem[\protect\citename{{American Psychological Association}}1983]{APA:83}
{American Psychological Association}.
\newblock 1983.
\newblock {\em Publications Manual}.
\newblock American Psychological Association, Washington, DC.

\bibitem[\protect\citename{{Association for Computing Machinery}}1983]{ACM:83}
{Association for Computing Machinery}.
\newblock 1983.
\newblock {\em Computing Reviews}, 24(11):503--512.

\bibitem[\protect\citename{Chandra \bgroup et al.\egroup }1981]{Chandra:81}
Ashok~K. Chandra, Dexter~C. Kozen, and Larry~J. Stockmeyer.
\newblock 1981.
\newblock Alternation.
\newblock {\em Journal of the Association for Computing Machinery},
  28(1):114--133.

\bibitem[\protect\citename{Gusfield}1997]{Gusfield:97}
Dan Gusfield.
\newblock 1997.
\newblock {\em Algorithms on Strings, Trees and Sequences}.
\newblock Cambridge University Press, Cambridge, UK.

\end{thebibliography}

\section{Supplementary Material}

\subsection{Approximate pseudo-likelihood}\label{pseudolikelihood} 
We compute the two autoregressive conditionals from forward and backward networks for each word $v_i$ 
 using respectively separate  position dependent hidden layers $\overrightarrow{\bf h}_i$ and $\overleftarrow{\bf h}_i$. To speed up computations, we can introduce a pseudo-likelihood
 
\begin{equation*}
p({\bf v}) \approx \prod_{i=1}^{D} p (v_i | {\bf v}_{\neg i})
\end{equation*}

\begin{align*}
\mbox{with }\, 
\begin{split}
p (v_i | {\bf v}_{\neg i}) & =  \frac{\exp (b_w + {\bf U}_{w,:} \overleftrightarrow{\bf h}_i ({\bf v}_{\neg i}))}{\sum_{w'} \exp (b_{w'} + {\bf U}_{w',:} \overleftrightarrow{\bf h}_i ({\bf v}_{\neg i}))}
\end{split}\\
\begin{split}
\overleftrightarrow{\bf h}_i({\bf v}_{\neg i}) & =  g (D{\bf c} +\sum_{k<i} {\bf W}_{:, v_k} + \sum_{k>i} {\bf W}_{:, v_k}) 
\end{split}
\end{align*}
Computing $\log p (v_i | {\bf v}_{\neg i})$ can speed up computation times by introducing a single hidden vector $\overleftrightarrow{\bf h}_i$ for each $i$ instead of using the full forward and backward conditionals.  
However, in our proposed formulation of iDocNADE (Figure \ref{fig:AutoregressiveNetworks}), we perform exact inference by computing $\mathcal{L}^{iDocNADE}({\bf v})$ as mean of the full forward and backward log likelihoods.


\begin{table}[t]
      \centering
        \begin{tabular}{c|c}
         \hline 
         {\bf Hyperparameter}               & {\bf Search Space} \\ \hline
           learning rate        &    [\underline{0.001}, 0.005, 0.01]                   \\
           hidden units        &       [50, 200]                      \\
           iterations        &      [2000]      \\
           activation function      &    sigmoid            \\ 
          scaling factor        &      [True, \underline{False}]  \\ \hline      
       \end{tabular}
         \caption{Hyperparameters in Generalization evaluation in the DocNADE and iDocNADE for 50 and 200 topics. The \underline{underline} signifies the optimal setting.}\label{HyperparametersinGeneralization}
\label{hyperparamtersppl}
  \end{table}%

\begin{table}[t]
      \centering
       \begin{tabular}{c|c}
        \hline 
        {\bf Hyperparameter}               & {\bf Search Space} \\ \hline
          retrieval fraction        &    [0.02]                        \\
          learning rate        &    [\underline{0.001}, 0.01]                       \\
          hidden units         &      [200]               \\ 
          activation function        &      [\underline{sigmoid}, {tanh}]      \\
          iterations        &      [2000, 3000]      \\
          scaling factor        &      [True, \underline{False}]            \\ \hline
         \end{tabular}

 \caption{Hyperparameters in the Document Retrieval task. The \underline{underline} signifies the optimal setting.}
\label{hyperparamtersIR}
\end{table}


\begin{table*}[t]
\centering
\renewcommand*{\arraystretch}{1.2}
\resizebox{0.9\textwidth}{!}{
\begin{tabular}{c}
\hline 

{\it Document Identifier:} {\bf T}    (from 20NewsGroups data set) \\ \hline

The CD-ROM and manuals for the March beta -- there is no X windows server there. \\
Will there be?  Of course.  (Even) if Microsoft supplies one with NT , other vendors will no doubt port their's to NT.\\ \hline \hline

{\it Document Identifier:} {\bf R}    (from Reuters21758 data set) \\ \hline 

SPAIN CARGO FIRMS HIRE DOCKERS TO OFFSET STRIKE  Cargo handling companies said they \\
were hiring twice the usual number of dockers to offset an  intermittent strike in Spanish ports. \\
Spanish dockers began a nine-day strike on Wednesday in  which they only work alternate hours \\
 in protest at government plans to partially privatize port services.\\ \hline \hline

\end{tabular}}
\caption{Raw text for the selected documents {\bf T} and {\bf R} from 20NewsGroups and Reuters21758 data sets, respectively.}
\label{rawtext}
\end{table*}

\end{document}